**Title**: Efficient vaccination strategies for epidemic control using network information

**Running Head**: Network measurement and vaccination strategy effectiveness

**Authors**: Yingrui Yang[†1], Ashley McKhann[†1], Guy Harling[2,3*], Jukka-Pekka Onnela[1*]

[†] These authors contributed equally to this work. [*] These authors contributed equally to this work.

**Author affiliations:**

1. Department of Biostatistics, Harvard T.H. Chan School of Public Health
2. Department of Epidemiology, Harvard T.H. Chan School of Public Health
3. Institute for Global Health, University College London

**Address for correspondence**: Jukka-Pekka Onnela. Department of Biostatistics, Harvard T.H. Chan School of Public Health, 655 Huntington Avenue, Boston, Massachusetts 02115, onnela@hsph.harvard.edu

**Word counts**:   Abstract:   300   Article:   4143

**Acknowledgements**: This research was supported by the National Institutes of Health grants P30-AG034420 (sub-award C14A11852), U54-GM088558, R01-AI112339 and R37-AI051164-12.

**Disclaimer**: The funders had no role in study design, data collection and analysis, decision to publish, or preparation of the manuscript.

**Authors' contributions**: JPO conceptualized the study. AM and YY conducted the analyses and summarized the results under supervision from GH and JPO. AM, YY and GH wrote the first paper draft. All authors contributed to study design, data interpretation and final revisions to the text.

**TITLE: Efficient vaccination strategies for epidemic control using network information**

**ABSTRACT**

**Background**: Social and contact networks affect both epidemic spread and intervention implementation. Network-based interventions are most powerful when the full network structure is known. However, in practice, resource constraints require decisions to be made based on partial network information. We investigated how the effectiveness of network-based vaccination schemes varied based on the accuracy of network data available at individual and village levels.

**Methods**: We simulated propagating a Susceptible-Infected-Recovered process on static empirical social networks from 75 rural Indian villages. First, we used regression analysis to predict the percentage of individuals ever infected (cumulative incidence) based on village-level network properties. Second, we simulated vaccinating 10% of each village at baseline, selecting vaccinees through one of five network-based approaches: random individuals (Random); random contacts of random individuals (Nomination); random high-degree individuals (High Degree); highest degree individuals (Highest Degree); or most central individuals (Central). The first three approaches require only sample data; the latter two require full network data. We also simulated imposing a limit on how many contacts an individual can nominate (Fixed Choice Design, FCD), which reduces the data collection burden but generates only partially observed networks.

**Results**: In regression analyses, we found mean and standard deviation of the degree distribution to strongly predict cumulative incidence. In simulations, the Nomination method reduced cumulative incidence by one-sixth compared to Random vaccination; full network methods reduced infection by two-thirds. The High Degree approach had intermediate effectiveness.

Somewhat surprisingly, FCD truncating individuals' degrees at three was as effective as using complete networks.

**Conclusions**: Using even partial network information to prioritize vaccines at either the village or individual level, i.e. determine the optimal order of communities or individuals within each village, substantially improved epidemic outcomes. Such approaches may be feasible and effective in outbreak settings, and full ascertainment of network structure may not be required.

**INTRODUCTION**

A signature characteristic of vaccination for the prevention of infectious disease outbreaks is the ability to exploit herd immunity. That is, not everyone in the population needs to receive a preventative intervention in order to substantially reduce epidemic severity. This saving of both time and resources that would otherwise have to be invested in vaccinating every person can be increased by careful targeting of vaccinations to maximize the effect of only immunizing a subset of the population. An extreme example of this is the ring vaccination approach taken to Smallpox elimination (Fenner et al., 1988), and adapted to a recent Ebola vaccine trial (Ebola ça Suffit Ring Vaccination Trial Consortium, 2015), where only those believed to be close contacts of current cases were offered the vaccine.

Various methods of targeting vaccine provision can be used to maximize the impact of vaccination when not all community members can be vaccinated at once, due to either cost or supply constraints. Common targeting approaches include focusing on populations either at highest risk of mortality if infected (e.g., the elderly and children) or at highest risk of transmitting to others at high mortality risk (e.g., healthcare workers and children)(Ajenjo et al., 2010; Bansal et al., 2006; Basta et al., 2009; Medlock and Galvani, 2009).

Individual-level social connections are another important predictor of acquisition and transmission risk, known prior to epidemic commencement (Christley et al., 2005). A considerable literature has arisen considering optimal methods for minimizing epidemic spread across networks. Common strategies include the targeting of highest-degree individuals (i.e., those with the most contacts (Eames et al., 2009)), those who are most central in a network (Holme et al., 2002), or those who act as bridges between different communities within a network (Chen et al., 2008). However, such methods often require enumeration of the entire social network, i.e. sociocentric data, in order to pinpoint the most important individuals. As a result, sociocentric approaches are typically both resource intensive to conduct and respondent intensive to complete, which reduces the feasibility of their application in real-world settings.

One proposed approach to reduce the cost of sociocentric data acquisition is to use fixed choice designs (FCD). An FCD is a network study design where the identified respondents are given a maximum number of contacts they can name; this reduces the time taken to conduct interviews and thus reduces both interview costs and the burden on respondents (McCarty et al., 2007). Past work has suggested that FCD affects several canonical network characteristics (Kossinets, 2006), and as a result affects predicted epidemic speed and cumulative incidence (Harling and Onnela, 2016); in both cases the nature of these effects depends on the structural properties of the underlying network. However, if FCD data approximately maintains the ordering or ranking of individuals on key measures, for example, the high-degree individuals are correctly identified as such even if degree estimates are biased, such an approach may provide an efficient halfway house between standard egocentric and sociocentric methods.

An alternative class of vaccination strategies does not try to make the best choices from full-network data, which is likely not available in most practical settings, but rather make better-than-random choices using less data. One such method is to vaccinate the friends of randomly chosen individuals, based on the fact that, on average, one's friends have more friends than one has (Feld, 1991). As well as being used in simulation studies (Cohen et al., 2003; Salathé and Jones, 2010), this approach has been used in empirical studies to detect an epidemic early in its course (Christakis and Fowler, 2010) and to improve take-up of a novel intervention (Kim et al., 2015).

An extension to this method uses random walks, i.e. interviewing an individual about all their friends, having them name one of their friends chosen at random, finding this new person and then repeating this process some number of times (Fernández-Gracia et al., 2017). This process generates a network sample from which individuals with specific network properties, e.g. locally central or locally bridging individuals, can be identified (Gong et al., 2013; Salathé and Jones, 2010).

Finally, another compromise approach might be to primarily use egocentric data, but in concert with some best-guess population-level metric. For example, if we have a rough estimate of the average number of relevant contacts, we could selectively vaccinate those with higher-than-average contact numbers. This approach would require more resources than random vaccination – since many interviewed individuals would be ineligible for vaccination– but fewer resources than conducting a sociocentric census – both in terms of reduced numbers of interviews, and a simpler set of survey questions.

Some of these approaches to vaccine deployment have previously been tested against one-another (Salathé and Jones, 2010; Thedchanamoorthy et al., 2014; Ventresca and Aleman, 2013). However, there is limited systematic evidence comparing a range of different intervention approaches requiring different levels of resource input, particularly using real-world or real-world-like (i.e. consistent with empirically observed) networks as opposed to archetypal or synthetic network structures. We therefore conducted simulations of epidemics on sets of empirical social networks from 75 villages in rural Karnataka, India, data for which were originally collected for a microfinance intervention {Banerjee, 2013 #2561}. We had two key goals: first, to predict the cumulative incidence of an epidemic in a village based on key network features of that village; and second, to identify the network-based vaccination scheme for each village that best minimized epidemic spread in that village.

**METHODS**

We built our approach on empirical social contact data collected from 75 villages in Karnataka, India as part of a microfinance intervention study in 2006 (Banerjee et al., 2013a, b). We defined a connection between two individuals (an undirected edge between two nodes $i$ and $j$) to exist if either $i$ or $j$ reported that the two of them had engaged in any of the 12 types of social interaction asked about in the study.

**Simulating a spreading process**

To simulate an epidemic, we ran a Susceptible-Infected-Recovered (SIR) process across each complete village network. We first selected 1% of nodes in each network to be infected uniformly at random to begin the SIR process, and these nodes represent the initially infected epidemic seed population. At each discrete time step, an infected node could infect at most one susceptible neighbor, i.e., we employed unit infectivity (Staples et al., 2015), under the assumption that a time step constitutes the smallest time unit required to infect at most one susceptible person. The SIR process used probability $\beta = 0.25$ for an infectious individual to infect a susceptible contact per time step, and probability $\gamma = 0.1$ for an infectious individual to recover to per time step. These values for $\beta$ and $\gamma$ lead to an approximate cumulative incidence of 40% of the population of a village in the absence of any intervention. These values were not chosen to replicate any particular epidemic, but rather to provide a level of infection that would allow the impact of different vaccination strategies to be seen.

**Network data collection methods**

As outlined above, there are a range of ways to collect data in order to measure network structure and the position of an individual within that network. For our study, we simulated three classes of approach. First, we used a fully-observed sociocentric network, corresponding to interviewing everyone and asking them to name all their contacts.

Collecting full network information is resource-intensive for both interviewers and respondents. A second, less data-intensive approach is Fixed Choice Design (FCD). In FCD, respondents are asked to name up to a maximum of $K$ contacts, limiting the number of contacts person $i$ can name to $k_i^{out} \leq K$, i.e., out-degree is truncated at $K$ for all nodes $i$. However, others can still nominate person $i$ as a contact. As a result, the observed number of contacts of $i$ (combining out-degree and in-degree nominations and treating them as symmetric or undirected edges), can be greater than $K$, and may in fact be the same as the person $i$'s true undirected degree ($k_i$) in the underlying fully-observed network. To simulate FCD, we first converted each undirected village network into a directed graph by replacing each undirected edge between a pair of contacts with two directed edges between them. We then rebuilt each network by randomly adding up to $K$ of each individual's outgoing edges to a new graph; if an individual had $k_i^{out} \leq K$ contacts, then all of their original out-edges were included. We then collapsed the truncated directed graph back to an undirected one, where we defined an edge to be present if a directed edge in either direction between the nodes was present. We truncated graphs using values for the threshold of $K = 1, \ldots, 10$.

Both full sociocentric and FCD methods require everyone in a village to be interviewed. A third approach is to use a sample of individuals to generate estimates of some network properties of interest, such as average degree. Such sampling can be random across the whole village or based on interviewing intensively within a few sub-groups within the population.

**Predicting village-level cumulative incidence**

Preliminary analysis suggested that using the $n = 75$ empirical villages alone resulted in insufficient statistical power to allow us to draw meaningful inference about village-level properties. We therefore used the Congruence Class Model (CCM) to generate a larger number of simulated networks that resembled the observed 75 networks based on the degree mixing matrix of the village networks (Goyal et al., 2014). The CCM is similar to the Exponential Random Graph Model (ERGM) (Hunter et al., 2008; Koskinen et al., 2013). However, unlike ERGM, CCM incorporates not only the point estimates of network statistics of interest, but also their variability, modelling posterior predictive distributions based on the probability distribution of specific networks properties.

The degree mixing matrix (DMM) for an undirected network is defined as the proportion of edges in the network that connect nodes of given degrees (Newman, 2003). For example, element (2,3) of this matrix corresponds to the proportion of edges in the network that connect nodes with degrees 2 to nodes with degree and 3. We estimated the DMM separately for each village. We then implemented a Markov chain Monte Carlo (MCMC) sampler using the Metropolis-Hastings algorithm to generate a collection of sample networks for each village, starting from the DMM of a randomly generated Erdős–Rényi (ER) network. The models were implemented using the CCMnet package in R (Goyal et al., 2014). To ensure MCMC convergence, we checked that the mean degree and DMM of model-generated networks were qualitatively similar to those for the empirical networks. We randomly selected 10 of the 75 empirical village networks for which the MCMC converged, and then drew 100 network samples for each from the posterior distribution of the DMM of each village network, resulting in a total of 1000 sampled networks. We then ran the SIR process 500 times on each model-generated network. For each SIR simulation, we recorded the cumulative incidence as the proportion of nodes ever infected. The village-level simulation approach is outlined in Figure 1.

For each of the 1000 generated networks, we calculated seven village-level network characteristics: mean degree; standard deviation of degree; network density; network size (number of nodes in the network; invariant within each empirical village); degree-assortativity (Newman, 2003); mean betweenness centrality; and the proportion of nodes in the largest connected component. We computed each characteristic first in the fully observed network, and then recomputed the same characteristics using different values for the out-degree truncation parameter $K$ to simulate FCDs with various threshold values.

To determine which network features were most useful in predicting village-level cumulative incidence, we ran linear regression models for the 500,000 simulated epidemics with each of the seven village characteristics obtained from the simulated networks in the form:

$$\text{CumulativeIncidence}_{ijk} = \beta_0 + \beta_c \times \text{NetworkCharacteristic}_{cjk} + \gamma \times \text{NetworkSize}_k + u_{ijk}$$

Here SIR simulations $i = 1, \ldots, 500$ are nested within model-generated networks $j = 1, \ldots, 100$ and empirical villages $k = 1, \ldots, 10$, and $c = 1, \ldots 6$. We compared the root mean squared error (RMSE) and Akaike Information Criterion (AIC) value of models containing none and all village characteristics with models containing every possible combination of one, two or three characteristics, to determine the most parsimonious set of predictors. AIC was obtained from a single regression model for each combination of predictor variables; RMSE was obtained using 10-fold cross-validation on the 1000 sample networks (Shao, 1993).

To obtain final RMSE and AIC estimates, we ran a three-level hierarchical mixed effects model of our preferred models in the form:

$$\text{CumulativeIncidence}_{ijk} \sim \beta_{0jk} + \beta_{cjk} \times \text{NetworkCharacteristic}_{cjk} + \gamma_k \times \text{NetworkSize}_k + u_{ijk}$$

$$\beta_{ajk} \sim \beta_{ak} + \epsilon_{ajk}$$

$$\beta_{ak} \sim \beta_a + \nu_{ak}$$

$$\gamma_k \sim \gamma_o + \mu_k$$

Where $a = \{0, c\}$ where again $c = 1, \ldots 6$. Here $\beta_{ajk}$ is the sample network-level effects for each network characteristic and $\beta_{ak}$ and $\gamma_k$ are the village-level effects for each network characteristic and village network size, respectively. In this model, $u_{ijk}, \epsilon_{ajk}, \nu_{ak}$ and $\mu_k$ are normally distributed random effects with mean zero, $\beta_{0jk}$ are random intercepts, and $\beta_{cjk}$ are random slopes. Our inference was focused on $\beta_c$ and $\gamma_0$.

Once we had arrived at a parsimonious set of characteristics from the full network models, we evaluated how much predictive power these same characteristics had for FCD network data. For each of the 1000 sample networks, we generated one FCD network at each truncation level and measured its characteristics to arrive at 1000 independent observations at each of 10 FCD levels of truncation. We then reran our preferred hierarchical regression model to obtain estimates of the RMSE and AIC value at each FCD level, predicting the full-network cumulative incidence from the characteristics of the FCD network. This enabled us to evaluate the extent of information gain when network features were based on the full networks compared to FCD-based truncated variants of those networks.

**Selecting individuals to vaccinate**

In our simulation, vaccination occurred prior to a disease outbreak, but we assumed vaccine availability to be limited, which led us to select which individuals to vaccinate before propagating an epidemic. We assumed that the vaccine was fully effective, and thus vaccinated individuals could never be infected, effectively removing them and their adjacent edges from the network. We conducted this analysis on all 75 empirical village networks. We considered six methods for selecting individuals for vaccination based on the methods outlined above. The first four of these do not require network information on all population members:

1) *None*. As a baseline or counterfactual scenario, we considered epidemics in which no village members were vaccinated.

2) *Random*. We randomly selected 10% of individuals from each village network for vaccination. This method represents a typical scenario where no network information is utilized, or the identities of the vaccinated individuals are uncorrelated with their network positions.

3) *Nomination*. We again randomly selected 10% of individuals in each network, and then simulated a process of having these individuals to nominate a friend at random to receive the vaccination. We required each nomination to be unique, so if $i$ and $j$ both nominated $k$, $j$ had to select someone else, so long as any of their contacts were unvaccinated; this ensured that approximately 10% of nodes were vaccinated.

4) *High degree*. We simulated interviewing individuals sequentially at random, asking them how many contacts they had (their degree, $k_i$, which we assumed they knew and reported without error) and vaccinating them only if their degree was sufficiently high. We implemented this by randomly selecting an individual in the network, and if their degree was greater than the median of all individuals pooled across the 75 villages (median: 6, interquartile range 4-11), we vaccinated them. We repeated this process until 10% of people in the village were vaccinated. On average, this implies interviewing 20% of the population, the same number as would have to be approached in the Nomination method. As a sensitivity analysis, we varied the degree cutoff value between 0 and 10. (Note that this is distinct from the threshold $K$ used in the context of FCD.) The High Degree approach requires prior knowledge or an estimate of the overall median (or other cutoff) degree; otherwise one would have to estimate that as part of the process, leading to some individuals being visited twice.

We also used two whole network methods for selecting individuals for vaccination. Within each method we varied the completeness of the network from FCD networks based on truncation at integer values $K = 0, 1, \ldots, 10$ to using data from the full non-truncated network:

5) *Highest degree.* We selected the 10% of individuals in each village with the highest degree, i.e., those with the most contacts. We identified these individuals based on the observable network, and thus when examining FCD networks, we based the node identification on only the truncated degree.

6) *Most central.* We selected the 10% of individuals in each village with the highest level of betweenness centrality: $c_B(v) = \sum_{s,t \in V} \frac{\sigma(s,t|v)}{\sigma(s,t)}$ (Brandes, 2001). Betweenness centrality is a global measure of individual *v*'s centrality in the network based on the proportion of shortest paths between all node pairs in the network that pass through individual *v*.

For each of the 75 empirical village networks, we simulated each method of selecting individuals for vaccination and ran the SIR process 500 times for each method at each level of the threshold for FCD (where applicable) in each village. We summarized the cumulative incidence seen across these 500 runs using 95% confidence intervals and compared across methods. The individual-level simulation approach is outlined in Figure 2.

**RESULTS**

The 75 Karnataka villages had between 354 and 1775 enumerated members (Table 1). Each village member was linked to a median of 6 others and connections were strongly degree-assortative (median $\rho = 0.33, IQR: 0.31 - 0.37$). In almost all villages, over 95% of individuals were part of the largest connected component. The 1000 simulated networks we generated from 10 of the Karnataka villages had similar size, mean degree and thus density to the empirical networks (Supplementary Table 1). Degree assortativity, the standard deviation of the degree distribution, and mean betweenness centrality were lower in the simulated networks, although aside from degree-assortativity, these values fell well within the empirically observed ranges.

**Predicting village-level cumulative incidence**

In these village-level analyses, we ran an SIR process across the 1000 simulated village networks; a mean of 66.2% (95%CI: 65.6%-66.7%) of individuals became infected in the epidemics. After running regression models containing all seven characteristics alone, and in all combinations of two or three, the model with the lowest RMSE contained two predictors, the mean degree and standard deviation of degree (Table 2 and Supplementary Table 2). This model had RMSE and AIC values lower than a model containing all seven predictors, and its RMSE was 1.3 percentage points, or 19%, lower than the null model containing only an intercept.

At each of the 10 levels of FCD degree truncation, we computed the mean and standard deviation of degree for each simulated network and ran a regression model using these two network features to predict cumulative incidence. Having full information about the contact network did not improve either predictive power (Figure 3) or model fit (Supplementary Figure 1) compared to FCD at truncation level $K = 3$.

**Selecting individuals to vaccinate**

In these individual-level analyses, we simulated vaccinating 10% of each village in advance of running the SIR process, and all intervention approaches significantly reduced cumulative incidence relative to no intervention (Figure 4). *Random* vaccination was the least effective vaccination approach, reducing cumulative incidence by 32.3% compared to no vaccination, while vaccinating a nominated friend (*Nomination*) reduced cumulative incidence by a *further* 10.7%. Vaccinating the first 10% of individuals interviewed with above-median degree (*High degree*) further improved effectiveness, leading to an average reduction in cumulative incidence compared to no vaccination of 48.2%. When we varied the *High degree* cutoff, any value greater or equal to six (the median degree) was significantly more effective than the Nomination method (Supplementary Figure 2).

Simulated vaccination methods based on full-network information – *Highest degree* and *Most central* – had very similar results and were markedly more effective than other approaches. At $K = 0$, these methods (and thus cumulative incidence) were equivalent to *Random* selection as expected, since no connections were ascertained. However, so long as degree truncation was no lower than 1, both methods outperformed *Nomination*; and for degree truncation $K \geq 3$, cumulative incidence was not meaningfully different from knowing the full network.

To account for the similarity of performance between *Highest degree* and *Most central* methods, we checked the correlation between degree and betweenness centrality rankings in the each of the 75 villages. The Pearson linear correlation ranged from 0.54 to 0.61 (mean of 0.56), suggesting a high but not collinear degree of similarity.

**DISCUSSION**

Using epidemic simulations on real-world and real-world-like social networks, we showed in this study that when ability to vaccinate an entire population is limited, using social contact network information can improve results compared to a random vaccination process at both the village and individual level.

At the village level, we provided evidence that communities with high mean degree and low degree variance, conditional on village size, are likely to have epidemics that infect a greater proportion of village members. Indeed, villages at the 5th of percentile of mean degree distribution in our simulation data had cumulative incidence 15 percentage points lower than those at the 95th percentile; the gap between the 5th and 95th percentiles of the variance of the degree distribution was almost 13 percentage points. Furthermore, we showed that these measures of village degree distribution were effectively captured by having respondents report in our simulation about their first (up to) three social contacts. While not as straightforward to measure as village size (i.e. number of individuals living in a village), the first and second moments of the degree distribution could potentially be evaluated from a sample of residents – reducing the overall interview burden – and since only truncated information is required, the interview burden on each individual could also be quite low.

At the individual level, we found that any approach that utilized network characteristics of individuals to selectively vaccinate 10% of the population led to a significant, and often substantial, reduction in cumulative incidence. Something as simple as vaccinating a randomly nominated social contact of randomly selected individuals reduced incidence by 4.4 percentage points, or 11% of the incidence rate seen if the randomly selected individuals themselves, rather than the individuals whom they nominated, were vaccinated.

A similar approach of only vaccinating randomly selected individuals if they had more than some minimum number of social contacts proved even more effective than the nomination approach once that minimum number was set at or above the median number of social contacts seen in the empirical data. Both of these methods, Nomination and High Degree with a cutoff at the median degree, would involve accessing 20% of the population and asking only a couple of questions to each individual.

Methods that incorporated information about an individual's network-wide position, rather than just how many people they were directly connected to, were even more effective, reducing cumulative incidence by two-thirds, compared to random vaccination. Even more impressively, these methods were almost as effective if the village-wide position of individuals was estimated not from the fully observed network, but instead from partially observed networks with degree truncation as low as K=3. Thus, even though the whole-network methods, Highest Degree and Most Central, would require information from all village members, this burden could be reduced to a small number of questions per person.

**Strengths and limitations**

Previous simulation and empirical studies have considered some of the methods we present above. However, we believe that this is the first study to directly compare all these approaches in a systematic way. By combining empirical data on social contacts within Indian villages with a series of simulation techniques, we have provided evidence on the relative usefulness of different network characteristics in targeting vaccination campaigns to maximize the efficiency of limited resources, as is likely to be the case in outbreaks of novel pathogens.

Our study also has some limitations however. First, our simulations are based on social contact data for specific rural villages in one state of India. While societies across the world are likely share some network characteristics (Apicella et al., 2012), this work could benefit from being tested in other populations; it is unclear to what extent our findings generalize to other settings. Second, we used an SIR infection process, which is overly simplistic for most infections. We additionally did not incorporate social distancing or other post-outbreak interventions that might have mitigated the infectious process, leading to very high estimated cumulative incidence rates. While this may mean that absolute effects were overestimated relative to real-world situations, we made the same assumptions in all our models, including traditional vaccination approaches, and consequently the strengths and weaknesses of different network-based approaches to vaccination relative to one another are valid.

**Conclusion**

We show that using network information to prioritize scarce vaccines at either the individual or village level substantially improved epidemic outcomes, even when networks were only partially observed, due to partial sampling of nodes, of edges, or of both. Such approaches may be feasible and effective in outbreak settings.

# REFERENCES

Ajenjo, M.C., Woeltje, K.F., Babcock, H.M., Gemeinhart, N., Jones, M., Fraser, V.J., 2010. Influenza vaccination among healthcare workers: ten-year experience of a large healthcare organization. Infect. Control Hosp. Epidemiol. 31(03): 233-240. doi: 10.1086/650449.

Apicella, C.L., Marlowe, F.W., Fowler, J.H., Christakis, N.A., 2012. Social networks and cooperation in hunter-gatherers. Nature 481(7382): 497-501. doi: 10.1038/nature10736.

Banerjee, A., Chandrasekhar, A.G., Duflo, E., Jackson, M.O., 2013a. The diffusion of microfinance. Science 341(6144): 1236498. doi: 10.1126/science.1236498.

Banerjee, A., Chandrasekhar, A.G., Duflo, E., Jackson, M.O., 2013b. The diffusion of microfinance, V9 ed. Abdul Latif Jameel Poverty Action Lab. Available at: http://hdl.handle.net/1902.1/21538

Bansal, S., Pourbohloul, B., Meyers, L.A., 2006. A comparative analysis of influenza vaccination programs. PLoS Med. 3(10): e387. doi: 10.1371/journal.pmed.0030387.

Basta, N.E., Chao, D.L., Halloran, M.E., Matrajt, L., Longini, I.M., 2009. Strategies for pandemic and seasonal influenza vaccination of schoolchildren in the United States. Am. J. Epidemiol. 170(6): 679-686. doi: 10.1093/aje/kwp237.

Brandes, U., 2001. A faster algorithm for betweenness centrality. Journal of Mathematical Sociology 25(2): 163-177. doi: 10.1080/0022250X.2001.9990249.

Chen, Y., Paul, G., Havlin, S., Liljeros, F., Stanley, H.E., 2008. Finding a better immunization strategy. Phys. Rev. Lett. 101(5): 058701. doi: 10.1103/PhysRevLett.101.058701.

Christakis, N.A., Fowler, J.H., 2010. Social network sensors for early detection of contagious outbreaks. PLoS ONE 5(9): e12948. doi: 10.1371/journal.pone.0012948.

Christley, R.M., Pinchbeck, G., Bowers, R., Clancy, D., French, N., Bennett, R., Turner, J., 2005. Infection in social networks: using network analysis to identify high-risk individuals. Am. J. Epidemiol. 162(10): 1024-1031. doi: 10.1093/aje/kwi308.

Cohen, R., Havlin, S., Ben-Avraham, D., 2003. Efficient immunization strategies for computer networks and populations. Phys. Rev. Lett. 91(24): 247901. doi: 10.1103/PhysRevLett.91.247901.

Eames, K.T., Read, J.M., Edmunds, W.J., 2009. Epidemic prediction and control in weighted networks. Epidemics 1(1): 70-76.

Ebola ça Suffit Ring Vaccination Trial Consortium, 2015. The ring vaccination trial: a novel cluster randomized controlled trial design to evaluate vaccine efficacy and effectiveness during outbreaks, with special reference to Ebola. Br. Med. J. 351: h3740. doi: 10.1136/bmj.h3740.

Feld, S.L., 1991. Why your friends have more friends than you do. Am. J. Sociol.: 1464-1477. doi: 10.1086/229693.

Fenner, F., Henderson, D.A., Arita, I., Jezek, Z., Ladnyi, I.D., 1988. Smallpox and its eradication. World Health Organization, Geneva.

Fernández-Gracia, J., Onnela, J.-P., Barnett, M.L., Eguíluz, V.M., Christakis, N.A., 2017. Influence of a patient transfer network of US inpatient facilities on the incidence of nosocomial infections. Scientific Reports 7(1): 2930. doi: 10.1038/s41598-017-02245-7.

Gong, K., Tang, M., Hui, P.M., Zhang, H.F., Younghae, D., Lai, Y.-C., 2013. An efficient immunization strategy for community networks. PLoS ONE 8(12): e83489. doi: 10.1371/journal.pone.0083489.

Goyal, R., Blitzstein, J., de Gruttola, V., 2014. Sampling networks from their posterior predictive distribution. Netw. Sci. 2(01): 107-131. doi: 10.1017/nws.2014.2.

Harling, G., Onnela, J.-P., 2016. Impact of degree truncation on the spread of a contagious process on networks. arXiv preprint arXiv:1602.03434.

Holme, P., Kim, B.J., Yoon, C.N., Han, S.K., 2002. Attack vulnerability of complex networks. Phys Rev E 65(5): 056109. doi: 10.1103/PhysRevE.65.056109.

Hunter, D.R., Handcock, M.S., Butts, C.T., Goodreau, S.M., Morris, M., 2008. ergm: A package to fit, simulate and diagnose exponential-family models for networks. Journal of Statistical Software 24(3): nihpa54860.

Kim, D.A., Hwong, A.R., Stafford, D., Hughes, D.A., O'Malley, J.A., Fowler, J.H., Christakis, N.A., 2015. Social network targeting to maximise population behaviour change: a cluster randomised controlled trial. Lancet 386(9989): 145-153. doi: 10.1016/S0140-6736(15)60095-2.

Koskinen, J.H., Robins, G.L., Wang, P., Pattison, P.E., 2013. Bayesian analysis for partially observed network data, missing ties, attributes and actors. Soc. Networks 35(4): 514-527. doi: 10.1016/j.socnet.2013.07.003.

Kossinets, G., 2006. Effects of missing data in social networks. Soc. Networks 28(3): 247-268. doi: 10.1016/j.socnet.2005.07.002.

McCarty, C., Killworth, P.D., Rennell, J., 2007. Impact of methods for reducing respondent burden on personal network structural measures. Soc. Networks 29(2): 300-315. doi: 10.1016/j.socnet.2006.12.005.

Medlock, J., Galvani, A.P., 2009. Optimizing influenza vaccine distribution. Science 325(5948): 1705-1708. doi: 10.1126/science.1175570.

Newman, M.E.J., 2003. Mixing patterns in networks. Phys Rev E 67(2): 1-13. doi: 10.1103/PhysRevE.67.026126.

Salathé, M., Jones, J.H., 2010. Dynamics and control of diseases in networks with community structure. PLoS Comput. Biol. 6(4): e1000736. doi: 10.1371/journal.pcbi.1000736.

Shao, J., 1993. Linear model selection by cross-validation. J. Am. Stat. Assoc. 88(422): 486-494.

Staples, P.C., Ogburn, E.L., Onnela, J.-P., 2015. Incorporating contact network structure in cluster randomized trials. Scientific Reports 3(5): 17581. doi: 10.1038/srep17581.

Thedchanamoorthy, G., Piraveenan, M., Uddin, S., Senanayake, U., 2014. Influence of vaccination strategies and topology on the herd immunity of complex networks. Social Network Analysis and Mining 4(1): 1-16. doi: 10.1007/s13278-014-0213-5.

Ventresca, M., Aleman, D., 2013. Evaluation of strategies to mitigate contagion spread using social network characteristics. Soc. Networks 35(1): 75-88. doi: 10.1016/j.socnet.2013.01.002.

**FIGURES AND TABLES**

**Figure 1: Flow diagram of the village-level study design**

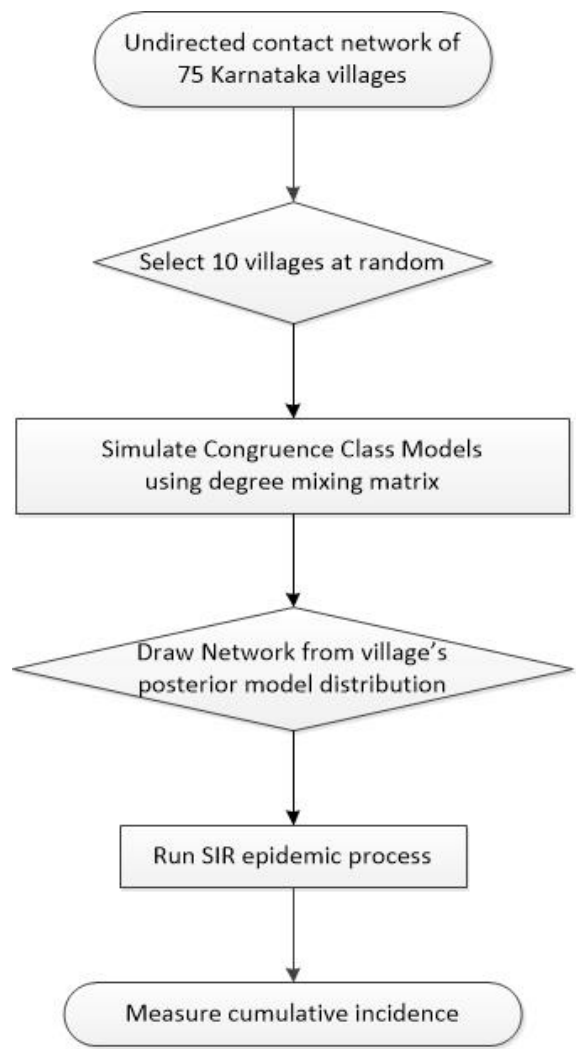

**Figure 2: Flow diagram of the individual-level study design**

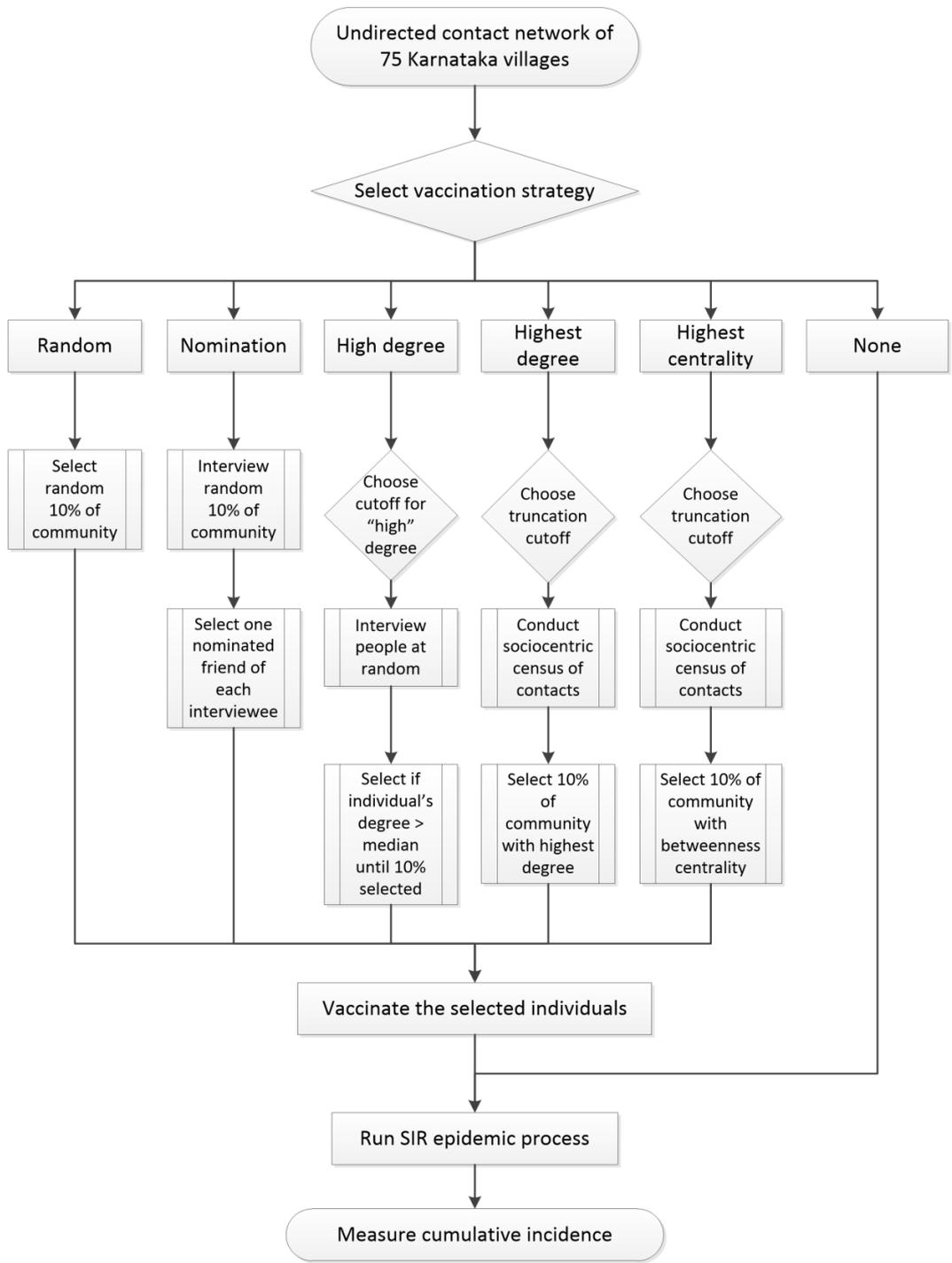

**Figure 3: Comparison of network characteristics to predict village-level cumulative incidence across different levels of network degree truncation using fixed choice design**

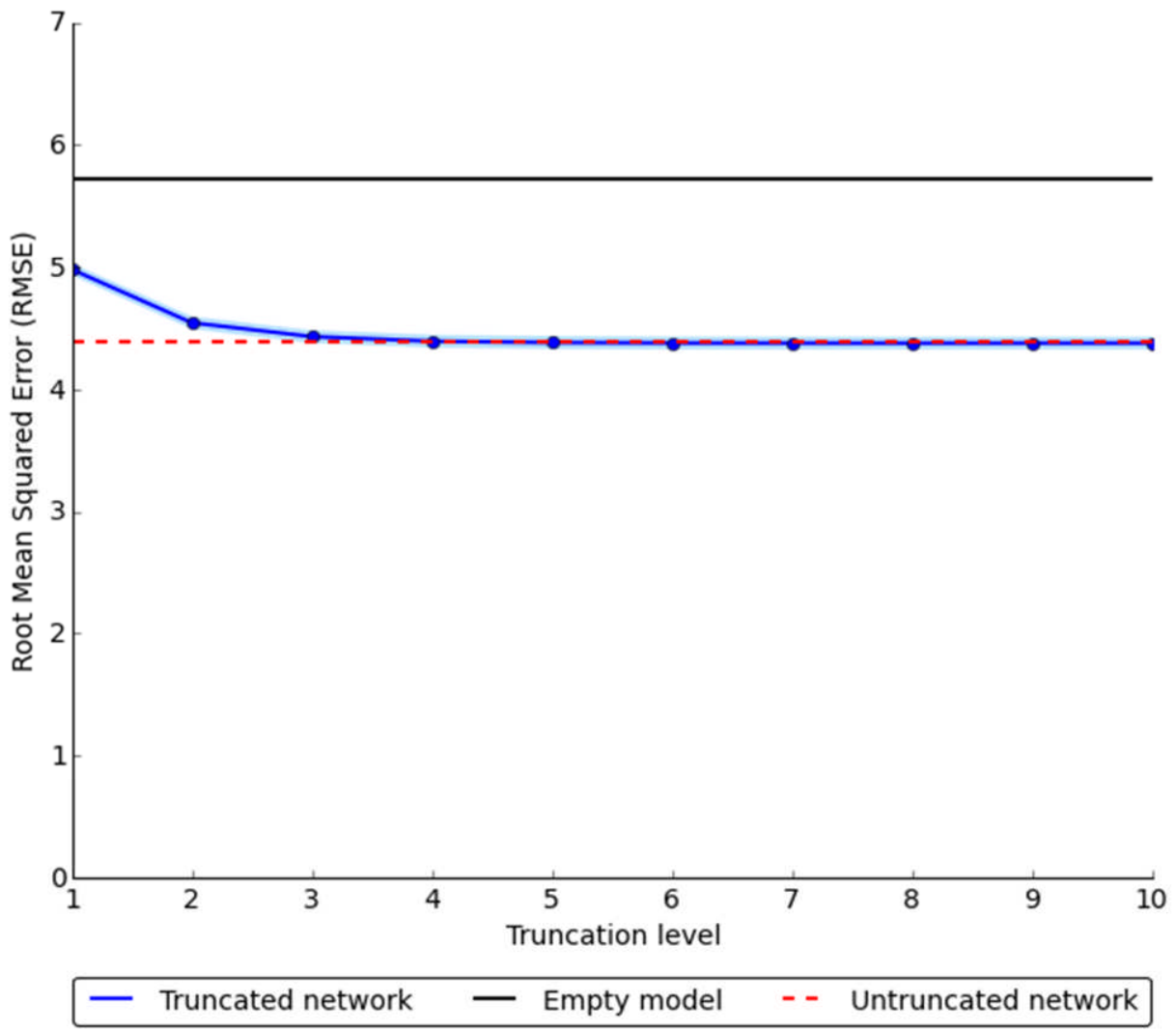

Numbers underlying this figure are provided in Supplementary Table 4. RMSE relates to cumulative incidence measured on (0-100) scale.

**Figure 4: Estimated cumulative incidence under different approaches to vaccinating 10% of each village**

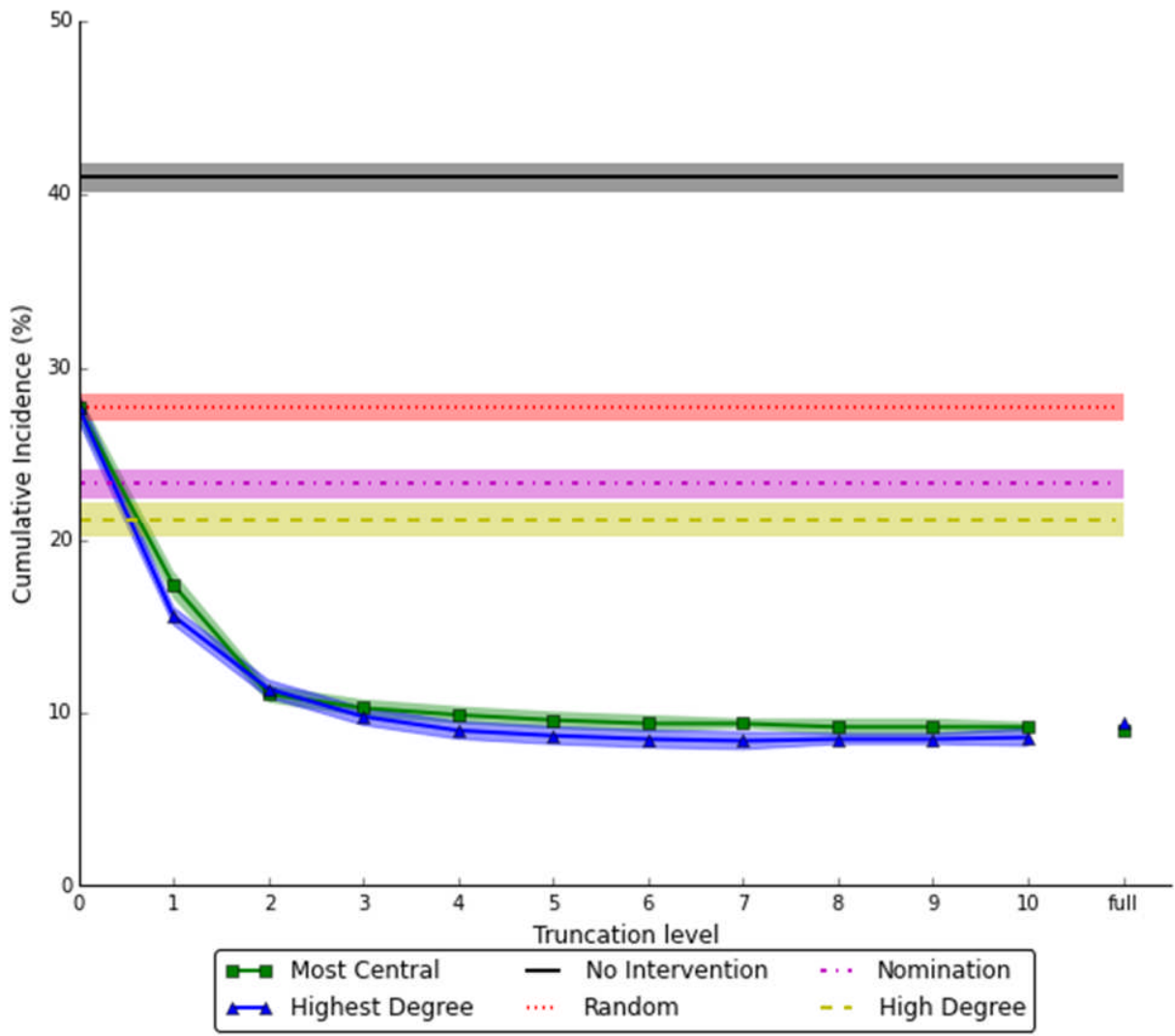

The six different vaccination methods are described in Figure 1. Solid or dashed lines and markers are point estimates; shaded areas represent 95% pointwise confidence intervals. Cumulative incidence is calculated as the mean of each of 75 villages' mean cumulative incidence across 500 SIR runs, i.e. $CI_{mean} = mean(mean(CI_j)_i)$, where $i$ indexes villages and $j$ indexes SIR runs. The confidence intervals are computed as $CI_{mean} \pm 1.96(SD(CI_i)/\sqrt{75})$, where SD is standard deviation. The High Degree method uses a cutoff of K=6, which corresponds to the median of the 75 village median degree values. Numbers underlying this figure are provided in Supplementary Table 3.

**Table 1: Characteristics of the full contact networks in 75 Karnataka villages**

|  | Median | Mean | 25% | 75% | Min | Max |
|---|---:|---:|---:|---:|---:|---:|
| Number of network members | 872.5 | 921 | 712 | 1140 | 354 | 1775 |
| Mean degree of network members | 8.4 | 8.5 | 7.8 | 9.0 | 6.8 | 10.4 |
| Median degree of network members | 6 | 6.41 | 6 | 7 | 5 | 8 |
| Standard deviation of degree | 5.8 | 6.0 | 5.2 | 6.5 | 9.8 | 8.7 |
| Network density (x$10^{-3}$) | 9.6 | 10.0 | 7.5 | 11.6 | 4.9 | 24.7 |
| Degree-assortativity | 0.33 | 0.34 | 0.31 | 0.37 | 0.15 | 0.53 |
| Mean betweenness centrality (x$10^{-3}$) | 3.3 | 3.5 | 2.7 | 4.1 | 1.9 | 6.7 |
| Percentage of nodes in the largest connected component | 97.4 | 96.9 | 96.3 | 98.3 | 88.7 | 99.9 |

All values for individual-level measures (i.e. the top five rows) are summary statistics of the relevant summary statistic from each of the 75 villages. All characteristics except median degree were included in models to predict village-level cumulative incidence.

**Table 2: Preferred predictive model of cumulative incidence using village-level characteristics**

|  | Empty model | Full model | | Model 1 | | Model 2 | |
|---|---|---|---|---|---|---|---|
| Mean degree | | 3.25 | [-3.14, 9.63] | 4.64 | [4.14 - 5.18] | 4.70 | [4.21 - 5.22] |
| Standard deviation of degree | | -4.05 | [-6.66, -1.44] | -3.95 | [-4.30 - -3.65] | -3.96 | [-4.29 - -3.64] |
| Number of network members | | -1.27 | [-14.6 , 12.0] | 0.27 | [-0.15 - 0.95] | | |
| Network density | | 1.24 | [-9.56, 12.0] | | | | |
| Degree-assortativity | | 0.23 | [-2.53, 2.99] | | | | |
| Mean betweenness centrality | | -3.11 | [-6.41, 0.19] | | | | |
| Percentage of nodes in the LCC | | 0.09 | [-1.93, 2.12] | | | | |
| Akaike information criterion (AIC) | 6323.4 | 5782.7 | | 5782.4 | | 5781.4 | |

The table presents regression coefficients and their 95% confidence intervals for the hierarchical three-level mixed-effects models for 500 SIR simulations on each of the 100 simulated networks from each of the selected 10 villages (total n=500,000). These 10 villages were chosen as explained in the text. Village-level characteristics were measured from empirical networks, although number of network members was invariant by design for networks simulated from any given village. Cumulative incidence is rescaled to percentage (0-100) of village population and village characteristics have been standardized, such that each regression coefficient represents the change in cumulative incidence in percentage points for a one-standard deviation change in the characteristic. For example, in Model 1, a one standard-deviation increase in mean degree is associated with a 4.64 percentage-point increase in cumulative incidence. LCC: largest connected component.

**SUPPLEMENTARY MATERIAL**

**Title**: Efficient vaccination strategies for epidemic control using network information

**Supplementary Figure 1: Comparison of the model fit for models of key network characteristics as predictors village-level cumulative incidence across levels of network data truncation**

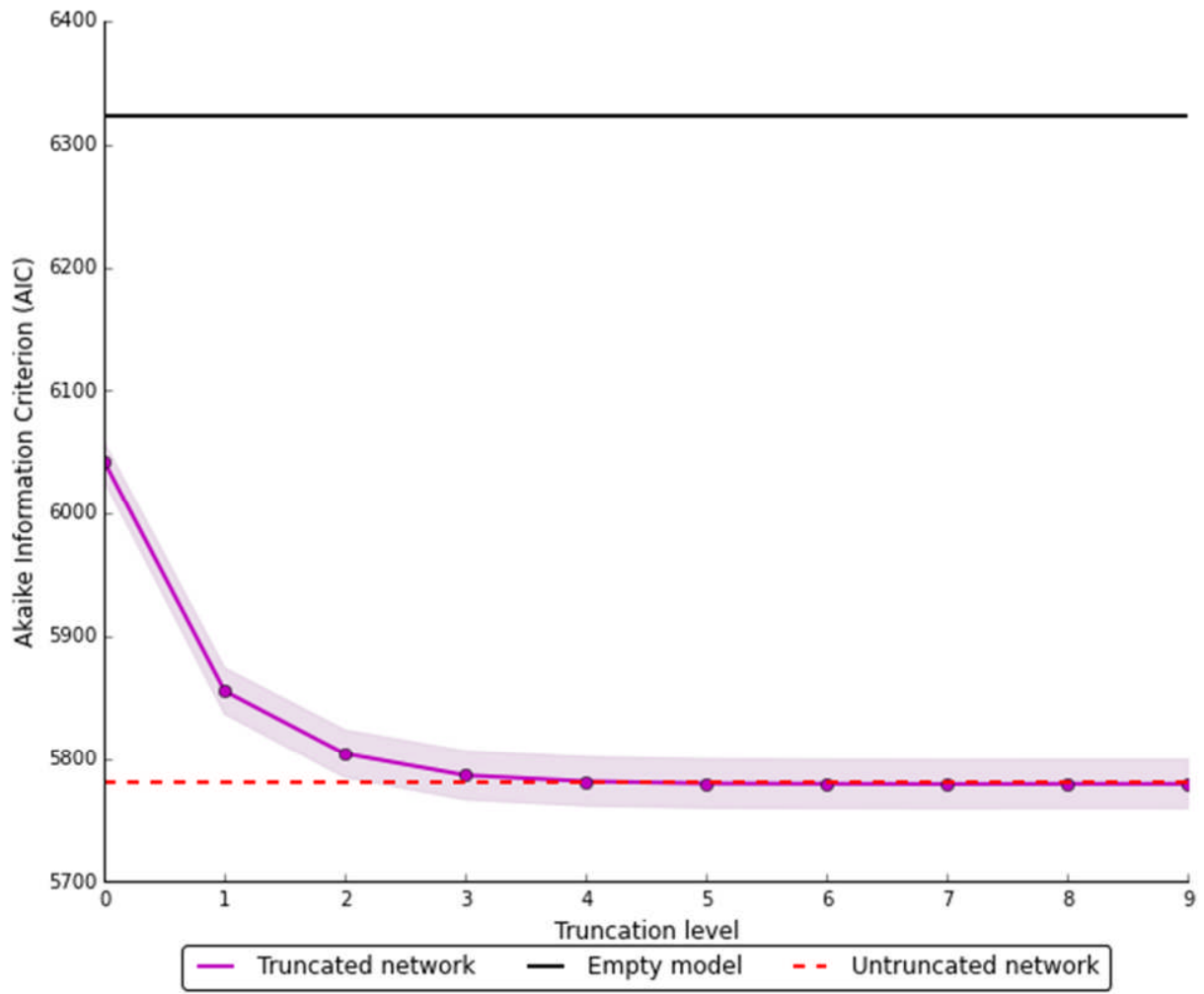

Numbers underlying this figure are provided in Supplementary Table 4. AIC: Akaike Information Criterion.

**Supplementary Figure 2: Estimated cumulative incidence under different approaches to vaccinating 10% of each village, with varying definition of a "high-degree" node**

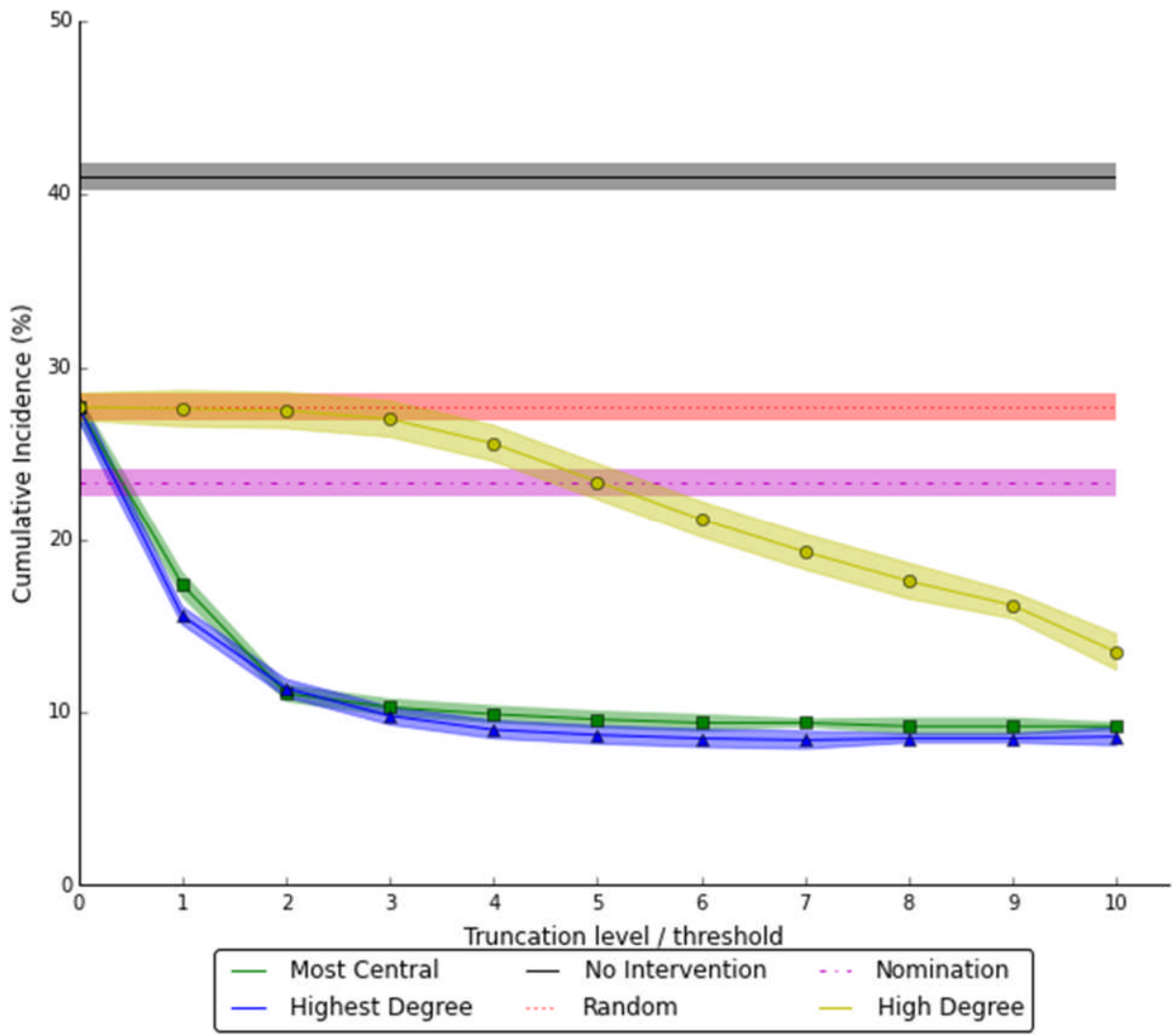

Cumulative incidence measured as percentage of the whole population. In this figure, the definition of "High Degree" varies by a cutoff value, where cutoff = 0 corresponds to 'Random', cutoff = 5 corresponds to choosing the first 10% of interviewed individuals (chosen at random) with a degree of 5 or greater. Numbers underlying this figure are provided in Supplementary Table 3.

**Supplementary Table 1: Characteristics of the 1000 simulated networks built from 10 empirical village networks**

|  | Median | Mean | 25% | 75% | Min | Max |
|---|---|---|---|---|---|---|
| Number of network members | 895.5 | 939.9 | 794 | 1025 | 650 | 1339 |
| Mean degree of network members | 8.3 | 8.2 | 7.2 | 9.0 | 6.9 | 9.5 |
| Standard deviation of degree | 4.4 | 4.5 | 4.1 | 4.9 | 3.6 | 5.8 |
| Network density ($\times 10^{-3}$) | 9.6 | 9.1 | 7.2 | 10.1 | 6.6 | 12.5 |
| Degree-assortativity | 0.11 | 0.11 | 0.08 | 0.13 | 0.02 | 0.21 |
| Mean betweenness centrality ($\times 10^{-3}$) | 2.7 | 2.7 | 2.6 | 3.0 | 1.8 | 3.7 |
| Percentage of nodes in the largest connected component | 100 | 99.9 | 99.8 | 100 | 98.4 | 100 |

**Supplementary Table 2: Summary of linear models predicting village-level cumulative incidence with full-network village-level characteristics**

| Model # | Density | Size | Mean degree | SD of degrees | Degree assortativity | LCC Proportion | Betweenness centrality | RMSE | AIC | AIC change |
|---|---|---|---|---|---|---|---|---|---|---|
| 1 | X | X | X | X | X | X | X | 4.39 | 5782.7 | |
| 2 | | | | | | | | 5.72 | 6323.4 | 540.7 |
| 3 | X | | | | | | | 5.72 | 6323.2 | 540.6 |
| 4 | | X | | | | | | 5.61 | 6284.9 | 502.2 |
| 5 | | | X | | | | | 5.12 | 6099.0 | 316.3 |
| 6 | | | | X | | | | 5.64 | 6291.8 | 509.1 |
| 7 | | | | | X | | | 5.72 | 6320.9 | 538.3 |
| 8 | | | | | | X | | 5.38 | 6198.4 | 415.7 |
| 9 | | | | | | | X | 5.55 | 6263.0 | 480.3 |
| 10 | X | X | | | | | | 5.15 | 6110.5 | 327.8 |
| 11 | X | | X | | | | | 5.11 | 6094.9 | 312.2 |
| 12 | X | | | X | | | | 5.64 | 6293.1 | 510.4 |
| 13 | X | | | | X | | | 5.71 | 6319.8 | 537.2 |
| 14 | X | | | | | X | | 5.37 | 6195.6 | 412.9 |
| 15 | X | | | | | | X | 5.23 | 6141.4 | 358.8 |
| 16 | | X | X | | | | | 5.11 | 6096.4 | 313.7 |
| 17 | | X | | X | | | | 5.38 | 6197.9 | 415.3 |
| 18 | | X | | | X | | | 5.62 | 6285.4 | 502.8 |
| 19 | | X | | | | X | | 5.32 | 6177.3 | 394.6 |
| 20 | | X | | | | | X | 5.52 | 6250.7 | 468.1 |
| 21 | | | X | X | | | | 4.38 | 5781.4 | -1.2 |
| 22 | | | X | | X | | | 5.11 | 6096.1 | 313.4 |
| 23 | | | X | | | X | | 4.88 | 6001.4 | 218.8 |
| 24 | | | X | | | | X | 5.09 | 6087.9 | 305.2 |
| 25 | | | | X | X | | | 5.63 | 6291.3 | 508.6 |
| 26 | | | | X | | X | | 5.38 | 6196.6 | 413.9 |
| 27 | | | | X | | | X | 5.18 | 6121.2 | 338.5 |
| 28 | | | | | X | X | | 5.38 | 6199.5 | 416.9 |
| 29 | | | | | X | | X | 5.55 | 6263.0 | 480.3 |
| 30 | | | | | | X | X | 5.24 | 6145.9 | 363.2 |
| 31 | X | X | X | | | | | 5.11 | 6094.6 | 311.9 |
| 32 | X | X | | X | | | | 4.55 | 5856.4 | 73.8 |
| 33 | X | X | | | X | | | 5.15 | 6110.1 | 327.5 |
| 34 | X | X | | | | X | | 4.95 | 6029.7 | 247.1 |
| 35 | X | X | | | | | X | 5.15 | 6110.8 | 328.2 |
| 36 | X | | X | X | | | | 4.38 | 5781.0 | -1.7 |
| 37 | X | | X | | X | | | 5.10 | 6090.2 | 307.6 |
| 38 | X | | X | | | X | | 4.86 | 5993.9 | 211.2 |
| 39 | X | | X | | | | X | 4.97 | 6035.9 | 253.3 |
| 40 | X | | | X | X | | | 5.64 | 6292.4 | 509.8 |
| 41 | X | | | X | | X | | 5.37 | 6195.1 | 412.5 |
| 42 | X | | | X | | | X | 4.46 | 5814.9 | 32.2 |
| 43 | X | | | | X | X | | 5.38 | 6196.4 | 413.7 |
| 44 | X | | | | X | | X | 5.22 | 6136.0 | 353.3 |
| 45 | X | | | | | X | X | 4.90 | 6010.3 | 227.7 |

| Model # | Density | Size | Mean degree | SD of degrees | Degree assortativity | LCC Proportion | Betweenness centrality | RMSE | AIC | AIC change |
|---|---|---|---|---|---|---|---|---|---|---|
| 46 | X | X | X | | | | | 4.38 | 5780.8 | -1.8 |
| 47 | X | X | | | X | | | 5.10 | 6091.4 | 308.8 |
| 48 | X | X | | | | X | | 4.86 | 5992.8 | 210.2 |
| 49 | X | X | | | | | X | 5.05 | 6069.7 | 287.0 |
| 50 | X | | X | X | | | | 5.38 | 6198.3 | 415.6 |
| 51 | X | | X | | | X | | 5.26 | 6153.4 | 370.7 |
| 52 | X | | X | | | | X | 5.01 | 6056.5 | 273.8 |
| 53 | X | | | X | X | | | 5.33 | 6178.5 | 395.8 |
| 54 | X | | | | X | | X | 5.51 | 6247.8 | 465.1 |
| 55 | X | | | | | X | X | 5.10 | 6089.5 | 306.9 |
| 56 | | X | X | X | | | | 4.38 | 5782.4 | -0.3 |
| 57 | | X | X | | | X | | 4.38 | 5781.3 | -1.3 |
| 58 | | X | X | | | | X | 4.38 | 5781.3 | -1.4 |
| 59 | | X | | X | X | | | 4.88 | 6002.2 | 219.5 |
| 60 | | X | | X | | | X | 5.08 | 6082.5 | 299.8 |
| 61 | | X | | | | X | X | 4.86 | 5991.6 | 209.0 |
| 62 | | | X | X | X | | | 5.38 | 6197.8 | 415.1 |
| 63 | | | X | X | | | X | 5.18 | 6121.1 | 338.5 |
| 64 | | | X | | | X | X | 5.09 | 6086.2 | 303.6 |
| 65 | | | | X | X | X | X | 5.25 | 6147.1 | 364.4 |

Each row in this table represents one linear regression model, where the outcome is cumulative incidence. The village-level predictors included in each model have been marked with an 'X'. The RMSE and AIC values are means across 500 simulations for each regression. AIC change is the difference in mean AIC for each model compared to the full model, Model #1, which contained all 7 predictors. Explanation of terms used: SD = standard deviation, RMSE = root mean squared error, AIC =

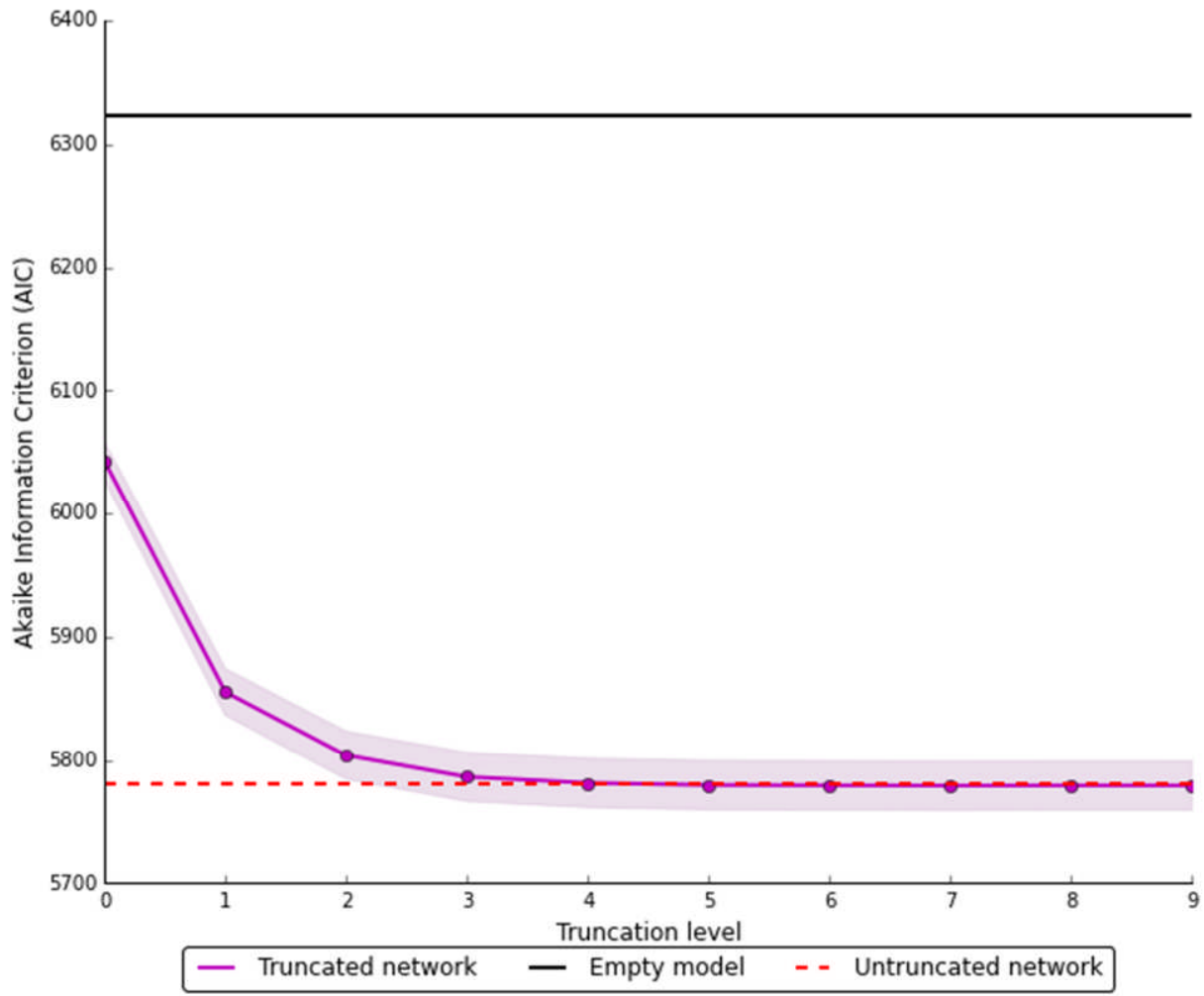

(AIC). Cumulative incidence is rescaled to percentage (0-100) of village population and village characteristics have been standardized, such that each regression coefficient represents the change in cumulative incidence in percentage points for a one-standard deviation change in the characteristic. For correspondence with Table 2: Model #1 here = Full model in Table 2, Model #2 here = Empty model in Table 2, Model #56 here = Preferred model 1 in Table 2, and Model #21 here = Preferred model 2 in Table 2.

**Supplementary Table 3: Summary of cumulative incidence for 500 SIR process runs on 75 Karnataka villages**

| Vaccination method | FCD level /cutoff | Mean | 95% CI | Min, max |
|---|---|---|---|---|
| None | | 41.0 | [40.2 - 41.8] | 26.9, 49.0 |
| Random | | 27.7 | [26.9 - 28.5] | 16.1, 35.2 |
| Nomination | | 23.3 | [22.5 - 24.1] | 14.0, 30.2 |
| High degree | 10 | 13.5 | [12.5 - 14.5] | 5.8, 22.2 |
| | 9 | 16.2 | [15.4 - 17.0] | 7.4, 25.6 |
| | 8 | 17.6 | [16.6 - 18.6] | 8.7, 27.0 |
| | 7 | 19.3 | [18.3 - 20.3] | 9.3, 28.3 |
| | 6 | 21.2 | [20.2 - 22.2] | 10.9, 29.4 |
| | 5 | 23.4 | [22.4 - 24.4] | 12.7, 32.0 |
| | 4 | 25.6 | [24.6 - 26.6] | 14.4, 33.2 |
| | 3 | 27.0 | [26.0 - 28.0] | 16.3, 34.7 |
| | 2 | 27.5 | [26.5 - 28.5] | 16.3, 35.1 |
| | 1 | 27.6 | [26.6 - 28.6] | 17.2, 35.2 |
| Most central | None | 9.0 | [8.5 - 9.5] | 5.6, 12.9 |
| | 10 | 9.2 | [8.9 - 9.5] | 5.8, 13.6 |
| | 9 | 9.2 | [8.7 - 9.7] | 5.7, 13.6 |
| | 8 | 9.2 | [8.7 - 9.7] | 5.8, 13.9 |
| | 7 | 9.4 | [9.1 - 9.7] | 5.9, 13.5 |
| | 6 | 9.4 | [8.9 - 9.9] | 6.1, 13.8 |
| | 5 | 9.6 | [9.1 - 10.1] | 5.9, 14.5 |
| | 4 | 9.9 | [9.4 - 10.4] | 6.2, 14.2 |
| | 3 | 10.3 | [9.8 - 10.8] | 6.1, 15.0 |
| | 2 | 11.1 | [10.6 - 11.6] | 6.7, 16.2 |
| | 1 | 17.4 | [16.6 - 18.2] | 8.8, 24.8 |
| Highest degree | None | 9.4 | [8.9 - 9.9] | 5.6, 17.2 |
| | 10 | 8.6 | [8.1 - 9.1] | 5.2, 14.2 |
| | 9 | 8.5 | [8.2 - 8.8] | 5.3, 13.0 |
| | 8 | 8.5 | [8.2 - 8.8] | 5.1, 12.4 |
| | 7 | 8.4 | [7.9 - 8.9] | 5.1, 12.1 |
| | 6 | 8.5 | [8.0 – 9.0] | 5.1, 12.5 |
| | 5 | 8.7 | [8.2 – 9.2] | 5.1, 12.3 |
| | 4 | 9.0 | [8.5 - 9.5] | 5.4, 13.0 |
| | 3 | 9.8 | [9.3 - 10.3] | 5.9, 14.0 |
| | 2 | 11.4 | [10.9 - 11.9] | 6.3, 15.7 |
| | 1 | 15.6 | [15.1 - 16.1] | 8.8, 21.1 |

Explanation of terms used: CI: Confidence Interval. FCD level: value of $K$ used when computing betweenness centrality (*Most central* method) and out-degree (*Highest degree* method). Cutoff: out-degree minimum value required to vaccinate in the *High degree* method; a value of 6 was used for the primary analysis in Figure 4. Mean and 95% CI are percentage points of cumulative infected individuals.

**Supplementary Table 4: Summary statistics for regression models using village-level network characteristics to predict village-level cumulative incidence across levels of network data truncation**

| Model | RMSE x $10^{-2}$ (SE) | AIC (SE) |
|---|---|---|
| Empty | 5.72 (0.42) | 6323.4 (143.2) |
| K=1 | 4.98 (0.46) | 6042.4 (179.8) |
| K=2 | 4.55 (0.50) | 5855.4 (213.9) |
| K=3 | 4.43 (0.50) | 5804.0 (222.7) |
| K=4 | 4.39 (0.51) | 5786.3 (226.7) |
| K=5 | 4.38 (0.51) | 5781.4 (228.0) |
| K=6 | 4.38 (0.51) | 5779.6 (228.1) |
| K=7 | 4.38 (0.51) | 5779.3 (227.7) |
| K=8 | 4.38 (0.51) | 5779.1 (227.8) |
| K=9 | 4.38 (0.51) | 5779.3 (227.8) |
| K=10 | 4.38 (0.51) | 5779.3 (227.7) |
| Full | 4.38 (0.51) | 5781.4 (227.7) |

Each row provides summary statistics from a single linear regression to predict cumulative incidence at the village level using Model 2 from Table 2. K denotes the level of truncation, i.e., all individuals' degrees were truncated at K. The standard error is evaluated across 500 simulations.

**Supplementary Table 5: Correlation between village characteristics**

|  | Density | Size | Mean degree | SD of degree | Degree assortativity | LCC proportion | Mean betweenness centrality |
|---|---|---|---|---|---|---|---|
| Density |  | -0.848 | 0.094 | 0.087 | -0.184 | 0.139 | 0.881 |
| Size |  |  | 0.238 | 0.202 | 0.287 | 0.031 | -0.893 |
| Mean degree |  |  |  | 0.852 | 0.160 | 0.465 | -0.310 |
| SD of degree |  |  |  |  | 0.102 | 0.173 | -0.316 |
| Degree assortativity |  |  |  |  |  | -0.091 | -0.162 |
| LCC proportion |  |  |  |  |  |  | 0.080 |
| Mean betweenness centrality |  |  |  |  |  |  |  |

Correlation of village-level network features across the 75 empirical village networks.